\documentstyle{elsart}

\begin{document}

\begin{frontmatter}
\hfill SISSA/100/95/EP

\title{The complete structure of the $WG_2$ algebra and its BRST
quantization  }
\author{Chuan-Jie Zhu}
\address{International School for Advanced Studies, Via Beirut 2--4,
I-34014 Trieste, Italy}
\address{Physics Department, Graduate School,
Chinese Academy of Sciences, P. O. Box 3908,
Beijing 100039, P. R. China}

\begin{abstract}
The complete structure of the $WG_2$ algebra is obtained from an
explicit realization by an abstract Virasoro algebra and a free
boson field. We then construct its BRST operator and find a
seven-parameter family of nilpotent BRST operators. These free
parameters are related to the canonical transformations of
the ghost, antighost fields which leave the total stress-energy
tensor and the antighost field $b$ invariant.
\end{abstract}

\end{frontmatter}

\section{Introduction}

Theres exist only three generic nonlinear extension of the Virasoro
algebra by a single spin-$s$ ($s>2$) field. These algebras are
related to the rank 2 Lie algebras. Two of them, $W_3$ and $WB_2$,
related to the $A_2$ and $B_2$ ($=C_2$) algebras, are
well-studied in the literature \cite{zam,review}. The other one,
$WG_2$, related to the $G_2$ algebra, is a very complicated
algebra. The structure of $WG_2$ is known by solving the bootstrap
constraints or the Jacobi identities \cite{boot,bonn,wat}.
No explicit relization
is known for $WG_2$. In application, it is important to know
some realizations of the algebra. On the other hand, the authors
of \cite{popea} starts directly from two free boson fields to
formulate string theoy with higher spin
extension of the Virasoro algebra. As emphasied in \cite{zhua,zhub},
this may give rise to results which are not true for abstract
$W$-algebras. So it is important to study things using
only the abstract
$W$-algebras. In this paper we will first derive the complete
structure of the $WG_2$ algebra. We use only an abstract
Virasoro algebra and a free boson field to construct an
explicit realization of $WG_2$. Because the original Virasoro
algebra and the boson field satisfy the Jacobi identities, the
consistency of the $WG_2$ algebra is automatically guaranteed.
Having the abstract $WG_2$ algebra in hand, we then forget
its realization and study its BRST quantization.
To simplify the construction of the BRST operator,
we require that $\{Q, b(w)\}$ gives the total
stress-energy tensor. Even with this restriction, the BRST
operator is not unique and has seven free parameters.  We will
show that these parameters are related to the canonical
transformation of the ghost, antighost fields which leave the
total stress-energy tensor and the antighost field $b(z)$ invariant.

\section{The $WG_2$ algebra}

Let's start from the basic fields.
$\phi$ is a free boson field.
$T_1$ is the stress-energy tensor for an abstract Virasoro
algebra. The basic OPEs are
\begin{eqnarray}
\partial \phi(z) \partial \phi(w) \, & \sim &
 - { 1 \over (z-w)^2 } ,
\\
T_1(z) T_1(w) \, & \sim & { \tilde{c}/2 \over (z-w)^4} +
\left( { 2 \over (z-w)^2 } + { 1  \over z-w } \partial_w\right)
T_1(w). \label{basic}
\end{eqnarray}
{}From these fields we can construct  a new stress-energy tensor
and a spin-6 field as follows:
\begin{eqnarray}
T & = & T_1 - {1 \over 2 } (\partial \phi)^2 + a_0
\partial^2 \phi, \\
W & = & x_1 (\partial \phi)^4 T_1 + x_2 (\partial \phi)^6 +
(\hbox{27 more terms}). \label{derived}
\end{eqnarray}

By using the basic OPEs of $T_1$ and $\partial \phi$, one then computes
the OPEs for $T$ and $W$. The requirement that  $T$ and $W$ form a
(nonlinear) $WG_2$
algebra fixes all the coefficient $x$'s.
The OPEs for the $WG_2$ algebra can then be derived.
The final result is\footnote{setting all derivatives of
primary and quasi primary fields to 0,
see the notation in \cite{zhuc}.}
\begin{eqnarray}
T(z) T(w) & \sim & { c/2 \over (z-w)^4 } + { 2 T(w) \over (z-w)^2 } +
\Lambda_1(w) + (z-w)^2 \Lambda_2(w) \nonumber \\
& + & (z-w)^4 \Lambda_4(w) +
(z-w)^6 \Lambda_7(w), \label{eq:opewa} \\
T(z) W(w) & \sim & { 6 \over (z-w)^2 } W(w) +
X_1(w) + (z-w) X_4(w) + (z-w)^2 X_2(w), \nonumber \\
& & \\
W(z) W(w) & \sim & { c/6 \over (z-ww)^{12} } +
{ 2 T(w) \over (z-w)^{10} } + { b_1 \Lambda_1(w) \over
(z-w)^8 }  \nonumber \\
& + & { 1\over (z-w)^6 } ( b_2 \Lambda_2(w) + b_3
\Lambda_3(w) + c_0 W(w) ) \nonumber \\
& + & { 1\over (z-w)^4 } (b_4 \Lambda_4(w)  +
b_5 \Lambda_5(w) + b_6 \Lambda_6(w) + c_1 X_1(w) ) \nonumber \\
& + & { 1 \over (z-w)^2 }
\left( \sum_{i=7}^{10} b_{i} \Lambda_i(w) +
c_2 X_2(w) + c_3 X_3(w) \right) , \label{eq:opeww}
\end{eqnarray}
where some quasi primary fields are defined in the above in terms
of the nonsingular terms in OPEs and the rest are
defined as follows:
\begin{eqnarray}
T(z) \Lambda_1(w) & \sim &  (\hbox{singular terms}) +
\Lambda_3(w) + (z-w)^2 \Lambda_5(w) \nonumber \\
& +  & (z-w)^3 \Lambda_{11}(w) +
(z-w)^4 \Lambda_8(w) , \\
T(z) \Lambda_3(w) & \sim &   (\hbox{singular terms})
+ \Lambda_6(w) + (z-w) \Lambda_{12}(w) + (z-w)^2 \Lambda_9(w),
\nonumber \\
& & \\
T(z) \Lambda_8(w) & \sim & (\hbox{singular terms})
+ \Lambda_{10}(w) , \\
T(z) X_1(w) & \sim & (\hbox{singular terms})
+ X_3(w) . \label{eq:quasidef}
\end{eqnarray}
Here $\Lambda_{11}$ and $\Lambda_{12} = -8\,\Lambda_{11}/5$
are spin-9 quasi
primary fields which don't appear in the OPEs of the
$WG_2$ algebra.
The various coefficients appearing in
$(\ref{eq:opeww})$ are given as follows:
\begin{eqnarray}
b_1 & = &  {{62}\over {22 + 5\,c}}, \qquad\qquad
b_2  =   {{-740 + 17\,c}\over
    {\left( -1 + 2\,c \right) \,\left( 68 + 7\,c \right) }},
\nonumber \\
b_3 & = &
  {{80\,\left( 35 + 139\,c \right) }\over
    {3\,\left( -1 + 2\,c \right) \,\left( 22 + 5\,c \right) \,
      \left( 68 + 7\,c \right) }},
\nonumber \\
b_4 & = &
  {{-6450504 - 5184658\,c + 17697\,{c^2} + 1630\,{c^3}}\over
    {25\,\left( -1 + 2\,c \right) \,\left( 46 + 3\,c \right) \,
      \left( 3 + 5\,c \right) \,\left( 68 + 7\,c \right) }},
\nonumber \\
b_5 & = &
  {{6310112 - 46423496\,c - 4678806\,{c^2} + 121395\,{c^3}}\over
    {15\,\left( -1 + 2\,c \right) \,\left( 46 + 3\,c \right) \,
      \left( 3 + 5\,c \right) \,\left( 22 + 5\,c \right) \,
      \left( 68 + 7\,c \right) }},
\nonumber \\
b_6 & = &
  {{8\,\left( -2179 + 57652\,c + 22992\,{c^2} \right) }\over
    {\left( -1 + 2\,c \right) \,\left( 46 + 3\,c \right) \,
      \left( 3 + 5\,c \right) \,\left( 22 + 5\,c \right) \,
      \left( 68 + 7\,c \right) }},
\nonumber \\
b_7 & = &
  {{-3422146368 - 1675894344\,c - 7636990\,{c^2} + 268425\,{c^3} +
      6625\,{c^4}}\over
    {25\,\left( -1 + 2\,c \right) \,\left( 46 + 3\,c \right) \,
      \left( 3 + 5\,c \right) \,\left( 68 + 7\,c \right) \,
      \left( 232 + 11\,c \right) }},
\nonumber \\
b_8 & = &
  {{2\left( 25916384896 - 45359110192\,c - 7010455500\,{c^2} +
        6482100\,{c^3} + 2541375\,{c^4} \right) }\over
    {225\,\left( -1 + 2\,c \right) \,\left( 46 + 3\,c \right) \,
      \left( 3 + 5\,c \right) \,\left( 22 + 5\,c \right) \,
      \left( 68 + 7\,c \right) \,\left( 232 + 11\,c \right) }},
\nonumber \\
b_9 & = &
  {{8\,\left( -1089457712 - 16862376526\,c - 1015716375\,{c^2} +
        28458300\,{c^3} \right) }\over
    {585\,\left( -1 + 2\,c \right) \,\left( 46 + 3\,c \right) \,
      \left( 3 + 5\,c \right) \,\left( 22 + 5\,c \right) \,
      \left( 68 + 7\,c \right) \,\left( 232 + 11\,c \right) }},
\nonumber \\
b_{10} & = &
  {{32\,\left( 15707 + 874936\,c + 172800\,{c^2} \right) }\over
    {\left( -1 + 2\,c \right) \,\left( 46 + 3\,c \right) \,
      \left( 3 + 5\,c \right) \,\left( 22 + 5\,c \right) \,
      \left( 68 + 7\,c \right) \,\left( 232 + 11\,c \right) }},
\nonumber \\
c_0^2 & = & { 400 (47 + 2\,c)^2(516 + 13\,c)^2(2+c)(c^2 - 388 c
+ 4) \over  3 (2 c -1)( 3c + 46)(3c + 286)(5c + 3 )(5c + 22)
(7c + 68)(11c+232) },
\nonumber \\
c_1 & = & {186 c_0 \over 13 c+ 516 c }, \qquad
c_2  =  { 9 (13 c- 694) c_0 \over 130 (c+2)(c+47) } ,\nonumber \\
c_3  & =&   { 12 (572 c + 2089) c_0 \over 5(c+2)(c+47)(516+13c) } .
\label{coeall}
\end{eqnarray}
We have checked that the above structure of the $WG_2$ algebra is
in agreement with that given in \cite{boot}.

\section{The BRST quantization of the $WG_2$ algebra}

Now we study the BRST quantization of the $WG_2$ algebra. Following
the standard procedure we introduce ghost, antighost paris
$(c(z), b(z))$ and  $(\delta(z), \alpha(z))$ for $T(z)$ and
$W(z)$ respectively. These ghost anti-ghost fields have spins
$(-1,2)$ and $(-5,6)$  and their mode expansions are as follows
\begin{eqnarray}
c(z)   &  = & \sum_n c_n z^{ -n + 1} , \qquad
b(z) =  \sum_n b_n z^{ -n - 2} , \\
\delta(z) & = & \sum_n \delta_n z^{ - n  + 5} , \qquad
\alpha(z) =  \sum_n \alpha_n z^{ - n  - 6} .  \label{mode}
\end{eqnarray}
These modes satisfy the usual anti-commutation relations which can be
derived from the following OPEs
\begin{eqnarray}
c(z) b(w) & \sim &   { 1\over z-w}, \\
\delta(z) \alpha(w)  & \sim & { 1\over z-w} . \label{modeb}
\end{eqnarray}
The other OPEs are all 0 (nonsingular).

Because of the complexity with normal ordering we will not use
mode expansions. All our calculation are done with (the holomorphic)
fields. The normal ordering for the ghost anti-ghost fields are such
that the following equations are true
\begin{eqnarray}
c(z) b(w) &=&  { 1 \over z-w } + : c(z) b(w) : , \label{modebc} \\
\delta(z) \alpha(w) & =& { 1\over z-w } + : \delta(z) \alpha(w) : .
\label{modead}
\end{eqnarray}
This is possible because all these fields are free fields.

With all the above knowledge, we now construct the quantum BRST
operator. One way to start is to construct the corresponding classical
BRST operator \cite{min}.
The quantum BRST operator is then assumed to be the same
form as the classical one with possible renormalization of some
coefficients and addition of some zero mode terms due to normal
ordering. By imposing the nilpotent condition, one would determine
all these coefficients. For linear algebras this route is
quite successful.
The same strategy has  been applied to $W_3$ \cite{min}
and in \cite{peter}
to a class of quadratic non-linear algebra. But the
simplicity of this construction doesn't apply to  more complicated
nonlinear algebras, such as $WB_2$ and $W_4$ \cite{zhub,horn}.

We will follow the same strategy used in \cite{zhub} for the
BRST quantization of the $WB_2$ and $W_4$ algebras.
The BRST operator is the contour integration of a spin-1
current $j(z)$ with ghost number 1. To simplify our calculations, we
require that the (anti-) commutator of the BRST operator $Q$
with the stress-energy tensor antighost $b(z)$ gives the
total stree-energy tensor:
\begin{eqnarray}
\{ Q, b(z) \}  =  T_{\rm tot} & \equiv &
T(z) + 2 c'(z)b(z) + c(z) b'(z) \nonumber \\
& + &  6 \delta'(z)\alpha(z) + 5 \delta(z)\alpha'(z)\, .
\end{eqnarray}
This fixes the dependence of the BRST current $j(z)$ on the ghost
field $c(z)$ to the the following form
\begin{equation}
j(z) = : c(z) \big( T(z) + c'(z)b(z) +
6 \delta'(z)\alpha(z) + 5 \delta(z)\alpha'(z)\big) :
+ \cdots .
\end{equation}
We will group the rest terms by their $(\delta,\alpha)$-ghost
number. By ghost number and spin counting, there are possible
terms with $(\delta, \alpha)$-ghost number from 1 to 5.
The ansatz for the ($(\delta, \alpha)$-)
ghost number 1 terms is
\begin{eqnarray}
j_1 & = & \delta\,(a W + m_1T^2 \delta' \alpha  + m_2 T'' \delta'\alpha
+ m_3 T' \delta'\alpha' + m_4 T\delta'\alpha'' \nonumber \\
& + & m_5 T\delta^{(3)} \alpha +m_6 \delta' \alpha^{(5)} +
m_7 \delta^{(3)}\alpha^{(3)} +  m_8 \delta^{(5)}\alpha' +
m_9 \delta''\delta'\alpha'\alpha ) .
\end{eqnarray}
Here $a$ is an arbitrary constant which set the normalization for
the $\delta$ ghost. It is set to be $a = \sqrt{2560504830}$.
$j_2$, $j_3$, $j_4$ and $j_5$
have 80, 124, 51 and 4 terms respectively.
We will not give their explicit
form here. After we have solved the nilpotent
condition, we will give a simplified form of the BRST operator in
an Appendix.
Writing the BRST operator as the sum of various $(\delta, \alpha)$
ghost number terms:
\begin{eqnarray}
Q & = & Q_0 + Q_1 + Q_2 + Q_3 + Q_4 + Q_5, \\
Q_i & = & \oint_{0}[\d z] j_i(z),
\end{eqnarray}
the nilpotent condtion $Q^2 = 0$ becomes
\begin{eqnarray}
0 & = & Q^2_0, \label{eqa} \\
0 & = & \{Q_0, Q_1\}, \label{eqb}  \\
0 & = & \{Q_0, Q_2\}  + Q_1^2, \label{eqc} \\
0 & = & \{Q_0, Q_3\} + \{Q_1, Q_2\}, \label{eqd} \\
0 & = & \{Q_0, Q_4\} + \{Q_1, Q_3\}  + Q_2^2, \label{eqe} \\
0 & = & \{Q_0, Q_5\} + \{Q_1, Q_4\}  + \{Q_2, Q_3\}, \label{eqf} \\
0 & = & \{Q_1, Q_5\} + \{Q_2, Q_4\}  + Q_3^2, \label{eqg} \\
0 & = & \{Q_2, Q_5\} + \{Q_3, Q_4\}, \label{eqh} \\
0 & = & \{Q_3, Q_5\}  + Q_4^2, \label{eqi} \\
0 & = & \{Q_4, Q_5\},  \label{eqj} \\
0 & = & Q^5_0, \label{eqk} \\
\end{eqnarray}
The first equation gives the critical central charge $c = 388$. The
second equation (\ref{eqb}) is satisfied only for $m_i = 0, i = 1,
\cdots, 9$. We then solve equations (\ref{eqc}) and
(\ref{eqd}) together to found a 7-parameter solution for
the 204 coefficients in $Q_2$ and $Q_3$. The real time consuming part
of the calculations is to check eqs. (\ref{eqe})--(\ref{eqg}).
The rest eqs. (\ref{eqh})--(\ref{eqk}) are satisfied automatically
by ghost, antighost counting. Because the calculations will
take too long a time by simply using the {\tt OPEdefs.m}
Mathematica package
\cite{thie}, one must write some other programmes to do the
calculation. Let us say a few words about how we actually
did the calculations.

First it is a good idea to split the bosonic and fermionic
part of every terms. Because we knew that only lower
spin ($\leq 11$) bosonic fields could appear in $Q^2$,
we can expand all the OPEs (including also nonsingular terms)
to a certain degree approparitely. The OPEs for quasi primary
fields are also needed. These can be obtained easily by using the
OPEs and the definitions for quasi primary fields
given in $(\ref{eq:opewa})$ to $(\ref{eq:quasidef})$.
The OPEs with the fermionic part
are not so easy. One can just try a simple example by computing
the OPEs of $B_i(z) = :c^{(i)}(z)\cdots\,c'(z)\,c(z)\,b^{(i)}(z)\,
\cdots \,b'(z)\,b(z):$ with itself. The time needed grows
quite rapidly with $i$.
Because all the fermionic ghost fields are free fields, we can use
eqs. (\ref{modebc}) and (\ref{modead}) to do all the possible
contractions
and then obtain the OPEs by expanding all fields around $w$.
Actually, most of the computing time are used in this Laurent
expansion. I have written a simple programme to do all these things
but only for these free fermionic ghosts. For the problem in
hand, the computations take much short time and it is possible to
finish all the calculation in about two weeks. Here is the result:
all the eqs. (\ref{eqe})--(\ref{eqg}) are satisfied and the
rest 55 coefficients in $Q_4$ and $Q_5$ are also found.
The complete solution
is quite long because there are seven free parameters.
After explaining the meaning of the 7 free parameters in Q and also
puting some terms to 0, I will give an explicit solution in
the Appendix.

\section{The canonical transformations of the ghost, antighost fields}

The canonical transformations of the ghost, antighost fields
discussed in \cite{horn,zhub} are similar transformations.
One easily verifies that the 3-parameter and 7-parameter canonical
transformations for $WB_2$ and $W_4$ ghost, antighost fields
are the following transformations:
\begin{equation}
(\hbox{field})  \longrightarrow \e^S \, (\hbox{field}) \, e^{-S}
 = \sum_{n = 0}^{\infty}{ 1 \over n!}
\underbrace{[S, \cdots, [S}_{ n \hbox{ times}},(\hbox{field})]\cdots],
\end{equation}
where $S$ is the the contour integration of a spin-1 current
of ghost number 0:
\begin{equation}
S = S_{WB_2} = \oint[\d z] \left( c_1 b'b\delta''\delta +
c_2 c\,b'\,b\,\delta + c_3\,b\,\delta'\,\delta\,\alpha\, \right) ,
\end{equation}
for $WB_2$ and
\begin{equation}
S = S_{WB_2} + \oint\,[\d z] \left(
c_4\,b\,\gamma'\,\beta\,\delta +
c_5\,b\,\gamma\,\beta'\,\delta +
c_6\,b\,\gamma\,\beta\,\delta' +
c_7\,b'\,b\,\gamma\,\beta\,\delta'\,\delta \right),
\end{equation}
for $W_4$ ghost, antighost fields. From these results, we learnt
that in order to obtain all the possible canonical transformations, one
simply writes down all the possible ghost number 0
spin-1 currents module total derivatives. Of course, we omit
some of the simplest canonical transformations generated
by the currents
$c\,b$, $\delta\,\alpha$ and $b\,\delta''$. These are just simple
rescaling and shifting of the ghost fields.

For $WG_2$ ghost, antighost field system, there are altogether 56
independent ghost number 0 spin-1 currents (including simple rescaling
and shifting). One thing to be noticed here is that these currents
can also involve the Virasoro field because the spin of the $\delta$
ghost is so low that there exists many ghost, antighost currents
with very low spin. The canonical transformation for $WG_2$ is a
56-parameter family transformations. It is not difficult to obtain
the full transformation group. The more interesting thing is the
subgroup of canonical transformations which leave the total
stress-energy tensor and the antighost $b(z)$ invariant.
This subgroup is a 10-parameter group. Nevertheless 3 parameters are
fixed by our ansatz for the BRST operator:
\begin{equation}
j(z) = c(z) T(z) + a\,\delta(z) W(z) +
(\hbox{3 or more ghost terms})\,.
\end{equation}
Simple rescaling of $c(z)$ and $\delta(z)$ are not allowed.
Also simple shifting of $c(z)$ by $\delta^{(4)}(z)$ is not allowed.
This then leaves a 7-parameter subgroup of canonical transformations
which leaves the total stress-energy tensor and the antighost
$b(z)$ invariant. This is why we obtain a seven-parameter family of
nilpotent BRST operator for $WG_2$ algebra.

Now we give the explicit form of the seven parameter generator
for the canonical
transformations. We split it into 2 groups according the number of
ghost antighost fields. We have
\begin{eqnarray}
J_1 & = & {c_1}\,T\,b''\,b'\,b\,\delta^{(3)}\,\delta''\,\delta +
{c_2}\,T\,b''\,b'\,b\,\delta^{(4)}\,\delta'\,\delta +
{c_3}\,T\,b^{(3)}\,b'\,b\,\delta^{(3)}\,\delta'\, \delta \nonumber \\
& + & {c_4}\,b''\,b'\,b\,\delta^{(3)}\,\delta''\,\delta'\,
  \delta\,\alpha
-  \left(6\,{c_1} + {{28\,{c_2}}\over 3} - 6\,{c_3} \right)
T^2\,b''\,b'\,b\,\delta''\,\delta'\,\delta
\nonumber \\
& - & \left({{2683\,{c_1}}\over {89}} + {{4508\,{c_2}}\over {89}} -
  {{2877\,{c_3}}\over {89}}\right)
T\,b^{(3)}\,b''\,b\,\delta''\,\delta'\,\delta \nonumber \\
& + & \left({{901\,{c_1}}\over {89}} + {{1521\,{c_2}}\over {89}} -
  {{894\,{c_3}}\over {89}} \right)
T\,b^{(4)}\,b'\,b\,\delta''\,\delta'\,\delta
\nonumber \\
& + & \left({{421687906\,{c_1}}\over {666165}} +
  {{745270691\,{c_2}}\over {666165}} -
  {{159173993\,{c_3}}\over {222055}}
- {{193925\,{c_4}}\over {13972}}\right)
b''\,b'\,b\,\delta^{(4)}\,\delta''\,\delta' \nonumber \\
& + & \left({{919625281\,{c_1}}\over {2664660}} +
  {{407256149\,{c_2}}\over {666165}} -
  {{1043606219\,{c_3}}\over {2664660}}
- {{207681\,{c_4}}\over {27944}}\right)
b''\,b'\,b\,\delta^{(4)}\,\delta^{(3)}\,\delta \nonumber \\
& + & \left({{48729017\,{c_1}}\over {222055}} +
  {{259636796\,{c_2}}\over {666165}} -
  {{167000729\,{c_3}}\over {666165}}
- {{219601\,{c_4}}\over {41916}}\right)
b''\,b'\,b\,\delta^{(5)}\,\delta''\,\delta \nonumber \\
& - & \left({{10612867\,{c_1}}\over {666165}}
+ {{5834259\,{c_2}}\over {222055}} -
  {{3663884\,{c_3}}\over {222055}}
+ {{5603\,{c_4}}\over {69860}}\right)
b''\,b'\,b\,\delta^{(6)}\,\delta'\,\delta \nonumber \\
& - & \left({{421687906\,{c_1}}\over {666165}} +
  {{745270691\,{c_2}}\over {666165}} -
  {{159173993\,{c_3}}\over {222055}}
- {{47109\,{c_4}}\over {3493}}\right)
b^{(4)}\,b'\,b\,\delta^{(3)}\,\delta''\,\delta \nonumber \\
& - & \left({{27535487\,{c_1}}\over {222055}} +
  {{98770929\,{c_2}}\over {444110}} -
  {{63720561\,{c_3}}\over {444110}}
- {{3447\,{c_4}}\over {998}}\right)
b^{(4)}\,b'\,b\,\delta^{(4)}\,\delta'\,\delta \nonumber \\
& + & \left({{2318203\,{c_1}}\over {7485}}
+ {{1370416\,{c_2}}\over {2495}} -
  {{879499\,{c_3}}\over {2495}} - {{49407\,{c_4}}\over {6986}}\right)
b^{(4)}\,b^{(3)}\,b\,\delta''\,\delta'\,\delta \nonumber \\
& - & \left({{2318203\,{c_1}}\over {9980}}
+ {{1027812\,{c_2}}\over {2495}} -
  {{2638497\,{c_3}}\over {9980}}
- {{148221\,{c_4}}\over {27944}}\right)
b^{(5)}\,b'\,b\,\delta^{(3)}\,\delta'\,\delta,
\end{eqnarray}
and
\begin{eqnarray}
J_2 & = & {c_5}\,T\,b'\,b\,\delta^{(3)}\,\delta'
+ {c_6}\,T\,b'\,b\,\delta^{(4)}\,\delta
+ {c_7}\,T\,b''\,b\,\delta''\,\delta' \nonumber \\
& +  & \left( {{356406389669\,{c_5}}\over {29792560608}} -
  {{1482203755439\,{c_6}}\over {148962803040}} +
  {{1385455753\,{c_7}}\over {9930853536}} \right)
b'\,b\,\delta^{(4)}\,\delta'' \nonumber \\
& + & \left({{111038273687\,{c_5}}\over {37240700760}} +
  {{74234639611\,{c_6}}\over {186203503800}} -
  {{981306029\,{c_7}}\over {12413566920}} \right)
b'\,b\,\delta^{(5)}\,\delta' \nonumber \\
& - & \left({{1261098246353\,{c_5}}\over {446888409120}} -
  {{3548929059731\,{c_6}}\over {2234442045600}} -
  {{19561388891\,{c_7}}\over {148962803040}} \right)
b'\,b\,\delta^{(6)}\,\delta \nonumber \\
& - & \left({{1406793619\,{c_5}}\over {164903472}} -
  {{1324530053\,{c_6}}\over {164903472}} -
  {{26925745\,{c_7}}\over {54967824}} \right)
b^{(3)}\,b\,\delta^{(3)}\,\delta' \nonumber \\
& + & \left({{1004852585\,{c_5}}\over {329806944}} -
  {{946092895\,{c_6}}\over {329806944}} -
  {{19232675\,{c_7}}\over {109935648}} \right)
b^{(3)}\,b\,\delta^{(4)}\,\delta \nonumber \\
& - & \left({{454062238\,{c_5}}\over {931017519}} +
  {{2526875414\,{c_6}}\over {4655087595}} +
  {{479486714\,{c_7}}\over {310339173}} \right)
b'\,b\,\delta^{(3)}\,\delta''\,\delta\,\alpha \nonumber \\
& - & \left({{2541405917\,{c_5}}\over {931017519}} +
  {{27844868791\,{c_6}}\over {4655087595}} +
  {{284378101\,{c_7}}\over {310339173}} \right)
b'\,b\,\delta^{(4)}\,\delta'\,\delta\,\alpha \nonumber \\
& - & \left({{7933248596\,{c_5}}\over {931017519}} +
  {{16502801672\,{c_6}}\over {931017519}} +
  {{1659334768\,{c_7}}\over {310339173}} \right)
b''\,b\,\delta^{(3)}\,\delta'\,\delta\,\alpha \nonumber \\
& - & \left({{15221893295\,{c_5}}\over {620678346}} +
  {{3102509207\,{c_6}}\over {620678346}} +
  {{1025874619\,{c_7}}\over {206892782}} \right)
b''\,b'\,\delta''\,\delta'\,\delta\,\alpha \nonumber \\
& - & \left({{5503700363\,{c_5}}\over {931017519}} +
  {{20969565827\,{c_6}}\over {931017519}} +
  {{1870488151\,{c_7}}\over {310339173}} \right)
b^{(3)}\,b\,\delta''\,\delta'\,\delta\,\alpha \nonumber \\
& + & \left({{4474\,{c_5}}\over {4289}}
+ {{60233\,{c_6}}\over {21445}} -
  {{100\,{c_7}}\over {4289}} \right)
T\,b''\,b\,\delta^{(3)}\,\delta  \nonumber \\
& + & \left({{523297\,{c_5}}\over {137248}}
+ {{417533\,{c_6}}\over {686240}} +
  {{34175\,{c_7}}\over {137248}} \right)
T\,b''\,b'\,\delta''\,\delta \nonumber \\
& + & \left({{125717\,{c_5}}\over {205872}}
+ {{3729697\,{c_6}}\over {1029360}} +
  {{29705\,{c_7}}\over {68624}} \right)
T\,b^{(3)}\,b\,\delta''\,\delta \nonumber \\
& + & \left({{529217\,{c_5}}\over {137248}}
+ {{423517\,{c_6}}\over {686240}} +
  {{30975\,{c_7}}\over {137248}} \right)
T\,b^{(3)}\,b'\,\delta'\,\delta \nonumber \\
& - & \left({{8959\,{c_5}}\over {68624}}
- {{770893\,{c_6}}\over {343120}} -
  {{15855\,{c_7}}\over {68624}} \right)
T\,b^{(4)}\,b\,\delta'\,\delta \nonumber \\
& + & \left({{1850\,{c_5}}\over {12867}}
+ {{374\,{c_6}}\over {12867}} -
  {{1000\,{c_7}}\over {12867}} \right)
T^2\,b'\,b\,\delta''\,\delta \nonumber \\
& + & \left({{1110\,{c_5}}\over {4289}}
+ {{1122\,{c_6}}\over {21445}} -
  {{600\,{c_7}}\over {4289}} \right)
T^2\,b''\,b\,\delta'\,\delta \nonumber \\
& - & \left({{9526\,{c_5}}\over {4289}}
+ {{34\,{c_6}}\over {4289}} -
  {{3990\,{c_7}}\over {4289}} \right)
T\,b'\,b\,\delta''\,\delta'\,\delta\,\alpha.
\end{eqnarray}

By using these expressions, I have checked explictly that
the 4 transformations associated with $J_1$
which change the $\delta$ ghost filed give exactly the
form fo the BRST operator obtained by explicit computations.
In particular one can use these canonical transformations
to put all $W$-dependent terms in $Q_3$ and $Q_4$ to 0. After
doing that, the BRST operator can be written as the sum of four
anticommuting nilpotnet operators:
\begin{equation}
Q = \tilde{Q}_0 + c_1 \tilde{Q}_1 + c_2 \tilde{Q}_2 +
c_3 \tilde{Q}_3, \qquad \{\tilde{Q}_i,\tilde{Q}_j\} = 0.
\end{equation}
The dependence of the BRST operator on the rest three free parameters
is linear. Also the rest canonical transformations associated
with $J_2$ are just
linear transformations. The explicit expression given in the Appendix
is $\tilde{Q}_0$, arranged by their $(\delta,\alpha)$ ghost number.
The other three $\tilde{Q}_i$'s are $\tilde{Q}_0$ exact.

\vfill\eject

\centerline{\bf Appendix: The explicit expression of the BRST current}

\nopagebreak
The BRST current of $\tilde{Q}_0$ contains various $(\delta,
\alpha)$-ghost number terms. The 80 ghost number 2 terms are
\nopagebreak
\begin{eqnarray}
\begin{array}{clclcl}
{{59673456869}\over {498960000}} &
b\delta^{(5)}\delta^{(4)} &
{{31195928179}\over {207900000}} &
b\delta^{(6)}\delta^{(3)} &
{{-3256883498089}\over {8731800000}} &
b\delta^{(7)}\delta''
\\
{{152213868677}\over {317520000}} &
b\delta^{(8)}\delta' &
{{-152213868677}\over {571536000}} &
b\delta^{(9)}\delta &
{{-5574877217}\over {81675000}} &
b\delta''\delta'\delta\alpha^{(5)}
\\
{{-5574877217}\over {24502500}} &
b\delta^{(3)}\delta'\delta\alpha^{(4)} &
{{-196813097}\over {1089000}} &
b\delta^{(3)}\delta''\delta\alpha^{(3)} &
{{26502125533}\over {32670000}} &
b\delta^{(3)}\delta''\delta'\alpha''
\\
{{-16985555249}\over {45738000}} &
b\delta^{(4)}\delta'\delta\alpha^{(3)} &
{{-932737741}\over {1663200}} &
b\delta^{(4)}\delta''\delta\alpha'' &
{{1328840539}\over {1089000}} &
b\delta^{(4)}\delta''\delta'\alpha'
\\
{{-52596629033}\over {54885600}} &
b\delta^{(4)}\delta^{(3)}\delta\alpha' &
{{863380219}\over {871200}} &
b\delta^{(4)}\delta^{(3)}\delta'\alpha &
{{-83896046221}\over {457380000}} &
b\delta^{(5)}\delta'\delta\alpha''
\\
{{-11539195369}\over {91476000}} &
b\delta^{(5)}\delta''\delta\alpha' &
{{-553073339}\over {1815000}} &
b\delta^{(5)}\delta''\delta'\alpha &
{{-195848181061}\over {274428000}} &
b\delta^{(5)}\delta^{(3)}\delta\alpha
\\
{{-20406600331}\over {245025000}} &
b\delta^{(6)}\delta'\delta\alpha' &
{{426766016729}\over {686070000}} &
b\delta^{(6)}\delta''\delta\alpha &
{{-6972875060203}\over {16008300000}} &
b\delta^{(7)}\delta'\delta\alpha
\\
{{19416854041}\over {16335000}} &
b\delta^{(3)}\delta''\delta'\delta\alpha'\alpha &
{{57029639797}\over {55440000}} &
Tb\delta^{(4)}\delta^{(3)} &
{{-12364869680501}\over {9147600000}} &
Tb\delta^{(5)}\delta''
\\
{{3220505762981}\over {9147600000}} &
Tb\delta^{(6)}\delta' &
{{-45278977634813}\over {192099600000}} &
Tb\delta^{(7)}\delta &
{{-69611369737}\over {52272000}} &
Tb'\delta^{(4)}\delta''
\\
{{80479622431}\over {1524600000}} &
Tb'\delta^{(5)}\delta' &
{{-7902513471619}\over {9147600000}} &
Tb'\delta^{(6)}\delta &
{{-710049471547}\over {152460000}} &
Tb''\delta^{(3)}\delta''
\\
{{4488333158933}\over {1829520000}} &
Tb''\delta^{(4)}\delta' &
{{-4807942668217}\over {1829520000}} &
Tb''\delta^{(5)}\delta &
{{48214065211}\over {34303500}} &
Tb^{(3)}\delta^{(3)}\delta'
\\
{{-5774383084717}\over {1829520000}} &
Tb^{(3)}\delta^{(4)}\delta &
{{459106098581}\over {228690000}} &
Tb^{(4)}\delta''\delta' &
{{-3727403771143}\over {1372140000}} &
Tb^{(4)}\delta^{(3)}\delta
\\
{{-2574566500831}\over {2286900000}} &
Tb^{(5)}\delta''\delta &
{{-432477750757}\over {1143450000}} &
Tb^{(6)}\delta'\delta &
{{97387696081}\over {65340000}} &
Tb\delta''\delta'\delta\alpha^{(3)}
\\
{{14580671833}\over {4356000}} &
Tb\delta^{(3)}\delta'\delta\alpha'' &
{{7730567377}\over {16940000}} &
Tb\delta^{(3)}\delta''\delta\alpha' &
{{-524120096617}\over {152460000}} &
Tb\delta^{(3)}\delta''\delta'\alpha
\\
{{84825821821}\over {22869000}} &
Tb\delta^{(4)}\delta'\delta\alpha' &
{{606431695753}\over {457380000}} &
Tb\delta^{(4)}\delta''\delta\alpha &
{{18452903729}\over {15246000}} &
Tb\delta^{(5)}\delta'\delta\alpha
\\
{{85278024871}\over {16940000}} &
Tb'\delta''\delta'\delta\alpha'' &
{{334254453767}\over {45738000}} &
Tb'\delta^{(3)}\delta'\delta\alpha' &
{{-107759818127}\over {457380000}} &
Tb'\delta^{(3)}\delta''\delta\alpha
\\
{{4751408701}\over {1089000}} &
Tb'\delta^{(4)}\delta'\delta\alpha &
{{23294487541}\over {4620000}} &
Tb''\delta''\delta'\delta\alpha' &
{{541682200891}\over {152460000}} &
Tb''\delta^{(3)}\delta'\delta\alpha
\\
{{718566565093}\over {457380000}} &
Tb^{(3)}\delta''\delta'\delta\alpha &
{{197\,a}\over {900}} &
Wb\delta''\delta' &
{{-9\,a}\over {100}} &
Wb\delta^{(3)}\delta
\\
{{-98\,a}\over {2475}} &
Wb'\delta''\delta &
{{-9\,a}\over {100}} &
Wb''\delta'\delta &
{{-4064194451083}\over {2744280000}} &
T^2b\delta^{(3)}\delta''
\\
{{2921472697793}\over {5488560000}} &
T^2b\delta^{(4)}\delta' &
{{-172159457821}\over {1097712000}} &
T^2b\delta^{(5)}\delta &
{{-46505121533}\over {1372140000}} &
T^2b'\delta^{(3)}\delta'
\\
{{-30539070601}\over {152460000}} &
T^2b'\delta^{(4)}\delta &
{{284730936659}\over {152460000}} &
T^2b''\delta''\delta' &
{{-994188795821}\over {1372140000}} &
T^2b''\delta^{(3)}\delta
\\
{{-214406574031}\over {2744280000}} &
T^2b^{(3)}\delta''\delta &
{{-118177330583}\over {784080000}} &
T^2b^{(4)}\delta'\delta &
{{183119385787}\over {228690000}} &
T^2b\delta''\delta'\delta\alpha'
\\
{{19456666073}\over {32670000}} &
T^2b\delta^{(3)}\delta'\delta\alpha &
{{5951595949}\over {5082000}} &
T^2b'\delta''\delta'\delta\alpha &
{{-103\,a}\over {1650}} &
TWb\delta'\delta
\\
{{407431932661}\over {457380000}} &
T''Tb\delta''\delta' &
{{-29420636167}\over {114345000}} &
T''Tb\delta^{(3)}\delta &
{{-12068938999}\over {114345000}} &
T''Tb'\delta''\delta
\\
{{-231200488649}\over {457380000}} &
T''Tb''\delta'\delta &
{{-77520792869}\over {914760000}} &
T^{(4)}Tb\delta'\delta &
{{1904078111}\over {5488560}} &
T^3b\delta''\delta'
\\
{{-9198368491}\over {68607000}} &
T^3b\delta^{(3)}\delta &
{{-8926920587}\over {137214000}} &
T^3b'\delta''\delta &
{{-28861261637}\over {171517500}} &
T^3b''\delta'\delta
\\
{{-12589653817}\over {114345000}} &
T''T^2b\delta'\delta &
{{-157515137}\over {6534000}} &
T^4b\delta'\delta .  &
&
\end{array}
\nonumber
\end{eqnarray}

There are 124 possible ghost number 3 terms. Some of them are
fixed by using the seven-parameter canonical transformations.
The rest 115 terms are as follows:
\begin{eqnarray}
\begin{array}{clcl}
{{-7934109042500138719113269}\over {1533791676186000000}} &
b'b\delta^{(5)}\delta^{(4)}\delta'' &
{{-1667368584022693256429372999}\over {253075626570690000000}} &
b'b\delta^{(6)}\delta^{(3)}\delta''
\\
{{-16666473279072934299304361}\over {2091534103890000000}} &
b'b\delta^{(6)}\delta^{(4)}\delta' &
{{-568989949468038155157773777}\over {379613439856035000000}} &
b'b\delta^{(6)}\delta^{(5)}\delta
\\
{{-3996922374558882014726837201}\over {590509795331610000000}} &
b'b\delta^{(7)}\delta^{(3)}\delta' &
{{-239950378585579439331129563}\over {84358542190230000000}} &
b'b\delta^{(7)}\delta^{(4)}\delta
\\
{{-1397471914359077916986248427}\over {708611754397932000000}} &
b'b\delta^{(8)}\delta''\delta' &
{{-177426073968190731181444427}\over {96628875599718000000}} &
b'b\delta^{(8)}\delta^{(3)}\delta
\\
{{-1971041306949715186374210191}\over {3543058771989660000000}} &
b'b\delta^{(9)}\delta''\delta &
{{-233732536262351277656915033}\over {3543058771989660000000}} &
b'b\delta^{(10)}\delta'\delta
\\
{{34024243484459847510641269}\over {30369075188482800000}} &
b^{(3)}b\delta^{(4)}\delta^{(3)}\delta'' &
{{162845556695764420711023713}\over {50615125314138000000}} &
b^{(3)}b\delta^{(5)}\delta^{(3)}\delta'
\\
{{14319989479529649797115247}\over {15184537594241400000}} &
b^{(3)}b\delta^{(5)}\delta^{(4)}\delta &
{{662170963749315819119972261}\over {379613439856035000000}} &
b^{(3)}b\delta^{(6)}\delta''\delta'
\\
{{1152799767141684025639421809}\over {759226879712070000000}} &
b^{(3)}b\delta^{(6)}\delta^{(3)}\delta &
{{616810504739088900353741563}\over {885764692997415000000}} &
b^{(3)}b\delta^{(7)}\delta''\delta
\\
{{579710950851494769940193353}\over {5314588157984490000000}} &
b^{(3)}b\delta^{(8)}\delta'\delta &
{{-84026779312214353332827993}\over {253075626570690000000}} &
b^{(5)}b\delta^{(4)}\delta''\delta'
\\
{{-2777065577483826084816919}\over {12653781328534500000}} &
b^{(5)}b\delta^{(4)}\delta^{(3)}\delta &
{{-60139791185852666093989963}\over {253075626570690000000}} &
b^{(5)}b\delta^{(5)}\delta''\delta
\\
{{-3559836953868589034092829}\over {66598849097550000000}} &
b^{(5)}b\delta^{(6)}\delta'\delta &
{{20916119887299280076956861}\over {664323519748061250000}} &
b^{(7)}b\delta^{(3)}\delta''\delta
\\
{{3736177184815092491406959}\over {295254897665805000000}} &
b^{(7)}b\delta^{(4)}\delta'\delta &
{{-1236099637605531047159317}\over {579773253598308000000}} &
b^{(9)}b\delta''\delta'\delta
\\
{{305543680943510417689229}\over {25307562657069000000}} &
b'b\delta^{(4)}\delta^{(3)}\delta''\delta'\alpha &
{{310663406860065680073211}\over {25307562657069000000}} &
b'b\delta^{(5)}\delta^{(3)}\delta''\delta\alpha
\\
{{119522985133420812681049}\over {12653781328534500000}} &
b'b\delta^{(5)}\delta^{(4)}\delta'\delta\alpha &
{{61343907267104214014431}\over {15817226660668125000}} &
b'b\delta^{(6)}\delta^{(3)}\delta'\delta\alpha
\\
{{-2713550593929745519201499}\over {126537813285345000000}} &
b'b\delta^{(7)}\delta''\delta'\delta\alpha &
{{618766950761853728954431}\over {12653781328534500000}} &
b''b\delta^{(4)}\delta^{(3)}\delta''\delta\alpha
\\
{{1522973603977375002414923}\over {25307562657069000000}} &
b''b\delta^{(5)}\delta^{(3)}\delta'\delta\alpha &
{{-168951016933549645912819}\over {3329942454877500000}} &
b''b\delta^{(6)}\delta''\delta'\delta\alpha
\\
{{967690718032038057905549}\over {8435854219023000000}} &
b''b'\delta^{(4)}\delta^{(3)}\delta'\delta\alpha &
{{5723656916482740608789}\over {200853671881500000}} &
b''b'\delta^{(5)}\delta''\delta'\delta\alpha
\\
{{447242299447329584285243}\over {3615366093867000000}} &
b^{(3)}b\delta^{(4)}\delta^{(3)}\delta'\delta\alpha &
{{-12258051644696011305541}\over {1150343757139500000}} &
b^{(3)}b\delta^{(5)}\delta''\delta'\delta\alpha
\\
{{133492916436566519985997}\over {665988490975500000}} &
b^{(3)}b'\delta^{(4)}\delta''\delta'\delta\alpha &
{{6023609758692629765467}\over {66951223960500000}} &
b^{(3)}b''\delta^{(3)}\delta''\delta'\delta\alpha
\\
{{333967635166384940431373}\over {5061512531413800000}} &
b^{(4)}b\delta^{(4)}\delta''\delta'\delta\alpha &
{{4259543507166022741129841}\over {25307562657069000000}} &
b^{(4)}b'\delta^{(3)}\delta''\delta'\delta\alpha
\\
{{3714661861828725086984107}\over {126537813285345000000}} &
b^{(5)}b\delta^{(3)}\delta''\delta'\delta\alpha &
{{13800716658064624117063}\over {4217927109511500000}} &
Tb'b\delta^{(4)}\delta^{(3)}\delta''
\\
{{213773474384983500107419}\over {50615125314138000000}} &
Tb'b\delta^{(5)}\delta^{(3)}\delta' &
{{-1020068015588230739429}\over {3163445332133625000}} &
Tb'b\delta^{(5)}\delta^{(4)}\delta
\\
{{130304558569038116783117}\over {23006875142790000000}} &
Tb'b\delta^{(6)}\delta''\delta' &
{{-359689583127547857672293}\over {253075626570690000000}} &
Tb'b\delta^{(6)}\delta^{(3)}\delta
\\
{{179010496901458732144219}\over {80524062999765000000}} &
Tb'b\delta^{(7)}\delta''\delta &
{{837309222531646562877127}\over {885764692997415000000}} &
Tb'b\delta^{(8)}\delta'\delta
\\
{{-5630710848055468467773}\over {8435854219023000000}} &
Tb''b\delta^{(4)}\delta^{(3)}\delta' &
{{-178356879627206203221467}\over {33743416876092000000}} &
Tb''b\delta^{(5)}\delta''\delta'
\\
{{270487127804498569503773}\over {101230250628276000000}} &
Tb''b\delta^{(5)}\delta^{(3)}\delta &
{{6022597969889820098814047}\over {506151253141380000000}} &
Tb''b\delta^{(6)}\delta''\delta
\\
{{8370798286109273004877913}\over {1181019590663220000000}} &
Tb''b\delta^{(7)}\delta'\delta &
{{-16862342598784223763053}\over {613516670474400000}} &
Tb''b'\delta^{(4)}\delta''\delta'
\end{array}
\nonumber
\end{eqnarray}

\begin{eqnarray}
\begin{array}{clcl}
{{78467717685240325145041}\over {6748683375218400000}} &
Tb''b'\delta^{(4)}\delta^{(3)}\delta &
{{70621091506892827835233}\over {4217927109511500000}} &
Tb''b'\delta^{(5)}\delta''\delta
\\
{{4513219642779261912142229}\over {168717084380460000000}} &
Tb''b'\delta^{(6)}\delta'\delta &
{{-1828019420509219629669029}\over {101230250628276000000}} &
Tb^{(3)}b\delta^{(4)}\delta''\delta'
\\
{{32154350223251801626211}\over {6748683375218400000}} &
Tb^{(3)}b\delta^{(4)}\delta^{(3)}\delta &
{{1143752821343429308313471}\over {50615125314138000000}} &
Tb^{(3)}b\delta^{(5)}\delta''\delta
\\
{{9689457743973834368694503}\over {506151253141380000000}} &
Tb^{(3)}b\delta^{(6)}\delta'\delta &
{{-126923326442664511063993}\over {2410244062578000000}} &
Tb^{(3)}b'\delta^{(3)}\delta''\delta'
\\
{{30647672433870628426459}\over {920275005711600000}} &
Tb^{(3)}b'\delta^{(4)}\delta''\delta &
{{1645184080821643125465299}\over {25307562657069000000}} &
Tb^{(3)}b'\delta^{(5)}\delta'\delta
\\
{{138827852960815631277769}\over {7230732187734000000}} &
Tb^{(3)}b''\delta^{(3)}\delta''\delta &
{{634702359450372316947259}\over {16871708438046000000}} &
Tb^{(3)}b''\delta^{(4)}\delta'\delta
\\
{{-47310996287475657635743}\over {1840550011423200000}} &
Tb^{(4)}b\delta^{(3)}\delta''\delta' &
{{107868521037574748926787}\over {5061512531413800000}} &
Tb^{(4)}b\delta^{(4)}\delta''\delta
\\
{{569591977630846094779621}\over {20246050125655200000}} &
Tb^{(4)}b\delta^{(5)}\delta'\delta &
{{8237565721119697090349}\over {809842005026208000}} &
Tb^{(4)}b'\delta^{(3)}\delta''\delta
\\
{{547949708460619385500073}\over {6748683375218400000}} &
Tb^{(4)}b'\delta^{(4)}\delta'\delta &
{{7591704229758292563127}\over {149970741671520000}} &
Tb^{(4)}b''\delta^{(3)}\delta'\delta
\\
{{316623867182060830615847}\over {20246050125655200000}} &
Tb^{(4)}b^{(3)}\delta''\delta'\delta &
{{901231495195899807974983}\over {101230250628276000000}} &
Tb^{(5)}b\delta^{(3)}\delta''\delta
\\
{{811764376867965846915707}\over {33743416876092000000}} &
Tb^{(5)}b\delta^{(4)}\delta'\delta &
{{501051706652339247509}\over {9918699846000000}} &
Tb^{(5)}b'\delta^{(3)}\delta'\delta
\\
{{794408731957024547027041}\over {33743416876092000000}} &
Tb^{(5)}b''\delta''\delta'\delta &
{{1240318467076214946524429}\over {101230250628276000000}} &
Tb^{(6)}b\delta^{(3)}\delta'\delta
\\
{{268246855030984122483697}\over {14461464375468000000}} &
Tb^{(6)}b'\delta''\delta'\delta &
{{763343717601600647355311}\over {236203918132644000000}} &
Tb^{(7)}b\delta''\delta'\delta
\\
{{-124466820073830843583}\over {1207537105500000}} &
Tb'b\delta^{(3)}\delta''\delta'\delta\alpha'' &
{{-811892247963042565787}\over {8452759738500000}} &
Tb'b\delta^{(4)}\delta''\delta'\delta\alpha'
\\
{{-5024769253737836707}\over {256144234500000}} &
Tb'b\delta^{(4)}\delta^{(3)}\delta'\delta\alpha &
{{-31042377258207200509}\over {939195526500000}} &
Tb'b\delta^{(5)}\delta''\delta'\delta\alpha
\\
{{-248609372040723083}\over {402512368500000}} &
Tb''b\delta^{(3)}\delta''\delta'\delta\alpha' &
{{-86351280390679118581}\over {2817586579500000}} &
Tb''b\delta^{(4)}\delta''\delta'\delta\alpha
\\
{{-1845656139030146677}\over {100628092125000}} &
Tb''b'\delta^{(3)}\delta''\delta'\delta\alpha &
{{-8654634665276756611}\over {201256184250000}} &
Tb^{(3)}b\delta^{(3)}\delta''\delta'\delta\alpha
\\
{{22420636199}\over {7387200}} &
T^2b'b\delta^{(5)}\delta''\delta &
{{91565220149}\over {203148000}} &
T^2b'b\delta^{(6)}\delta'\delta
\\
{{1108880152796373919}\over {270488311632000}} &
T^2b''b\delta^{(4)}\delta''\delta &
{{815840987800876999}\over {270488311632000}} &
T^2b''b\delta^{(5)}\delta'\delta
\\
{{-100512367423}\over {40629600}} &
T^2b''b'\delta^{(3)}\delta''\delta &
{{157604654378837653}\over {30054256848000}} &
T^2b''b'\delta^{(4)}\delta'\delta
\\
{{78084563531401661}\over {19320593688000}} &
T^2b^{(3)}b\delta^{(3)}\delta''\delta &
{{983367807774000883}\over {270488311632000}} &
T^2b^{(3)}b\delta^{(4)}\delta'\delta
\\
{{31473963893287721}\over {6440197896000}} &
T^2b^{(3)}b'\delta^{(3)}\delta'\delta &
{{-4770998072911889}\over {1610049474000}} &
T^2b^{(3)}b''\delta''\delta'\delta
\\
{{1017236448542048501}\over {270488311632000}} &
T^2b^{(4)}b\delta^{(3)}\delta'\delta &
{{523706053025959297}\over {90162770544000}} &
T^2b^{(4)}b'\delta''\delta'\delta
\\
{{386807744976622463}\over {270488311632000}} &
T^2b^{(5)}b\delta''\delta'\delta &
{{742729403}\over {677160}} &
T^2b'b\delta^{(3)}\delta''\delta'\delta\alpha
\\
{{6662616373566101}\over {19320593688000}} &
T''Tb'b\delta^{(3)}\delta''\delta &
{{-93154238427865849}\over {22540692636000}} &
T''Tb'b\delta^{(4)}\delta'\delta
\\
{{-1552126807894351}\over {1073366316000}} &
T''Tb''b\delta^{(3)}\delta'\delta &
{{52065275724588347}\over {6440197896000}} &
T''Tb''b'\delta''\delta'\delta
\\
{{30060539578788199}\over {19320593688000}} &
T''Tb^{(3)}b\delta''\delta'\delta &
{{144086319689721929}\over {67622077908000}} &
T^{(4)}Tb'b\delta''\delta'\delta
\\
{{1190621242967}\over {1940598000}} &
T^3b'b\delta^{(3)}\delta''\delta &
{{-83824942877}\over {485149500}} &
T^3b'b\delta^{(4)}\delta'\delta
\\
{{471463088837}\over {4608920250}} &
T^3b''b\delta^{(3)}\delta'\delta &
{{4544925946691}\over {12290454000}} &
T^3b''b'\delta''\delta'\delta
\\
{{-8617452102727}\over {36871362000}} &
T^3b^{(3)}b\delta''\delta'\delta &
{{-88001317417}\over {62073000}} &
T''T^2b'b\delta''\delta'\delta
\\
{{-49278745789}\over {646866000}} &
T^4b'b\delta''\delta'\delta .  &
&
\end{array}
\nonumber
\end{eqnarray}

There are 51 ghost number 4 terms. Only one term depending on
$W$ is set to zero. The rest 50 terms are as follows:
\begin{eqnarray}
\begin{array}{clcl}
{{894012203254112689160479}\over {268966882800000000}} &
b''b'b\delta^{(5)}\delta^{(4)}\delta''\delta' &
{{401944511068676207595847}\over {427901859000000000}} &
b''b'b\delta^{(5)}\delta^{(4)}\delta^{(3)}\delta
\\
{{40752983873787784781775541}\over {14120761347000000000}} &
b''b'b\delta^{(6)}\delta^{(3)}\delta''\delta' &
{{28258929465723947578300037}\over {7060380673500000000}} &
b''b'b\delta^{(6)}\delta^{(4)}\delta''\delta
\\
{{3976133177714728853439929}\over {3137946966000000000}} &
b''b'b\delta^{(6)}\delta^{(5)}\delta'\delta &
{{136545858316535273628988667}\over {49422664714500000000}} &
b''b'b\delta^{(7)}\delta^{(3)}\delta''\delta
\\
{{10172277510601489892564111}\over {4118555392875000000}} &
b''b'b\delta^{(7)}\delta^{(4)}\delta'\delta &
{{43376645108126177448410147}\over {26358754514400000000}} &
b''b'b\delta^{(8)}\delta^{(3)}\delta'\delta
\\
{{88568028449146814148822683}\over {197690658858000000000}} &
b''b'b\delta^{(9)}\delta''\delta'\delta &
{{-2495489400013616746184393}\over {1255178786400000000}} &
b^{(4)}b'b\delta^{(4)}\delta^{(3)}\delta''\delta'
\\
{{-40920180369199307232545177}\over {18827681796000000000}} &
b^{(4)}b'b\delta^{(5)}\delta^{(3)}\delta''\delta &
{{-32673342484643079973559}\over {24515210671875000}} &
b^{(4)}b'b\delta^{(5)}\delta^{(4)}\delta'\delta
\\
{{-6168228851090077664487283}\over {2689668828000000000}} &
b^{(4)}b'b\delta^{(6)}\delta^{(3)}\delta'\delta &
{{-50242483093204085072693629}\over {43931257524000000000}} &
b^{(4)}b'b\delta^{(7)}\delta''\delta'\delta
\\
{{-2418408258415377900240491}\over {2091964644000000000}} &
b^{(5)}b'b\delta^{(4)}\delta^{(3)}\delta''\delta &
{{-1472100886651964423431343}\over {871651935000000000}} &
b^{(5)}b'b\delta^{(5)}\delta^{(3)}\delta'\delta
\\
{{-12609311056293005329010411}\over {10459823220000000000}} &
b^{(5)}b'b\delta^{(6)}\delta''\delta'\delta &
{{-16794859325642088061703839}\over {18827681796000000000}} &
b^{(6)}b'b\delta^{(4)}\delta^{(3)}\delta'\delta
\\
{{-10342168599291247263198641}\over {10459823220000000000}} &
b^{(6)}b'b\delta^{(5)}\delta''\delta'\delta &
{{26532493353233040719}\over {92217216960000}} &
b^{(6)}b^{(3)}b\delta^{(3)}\delta''\delta'\delta
\\
{{-73310271906663269153}\over {95632669440000}} &
b^{(7)}b'b\delta^{(4)}\delta''\delta'\delta &
{{-11202877685146493099}\over {40985429760000}} &
b^{(8)}b'b\delta^{(3)}\delta''\delta'\delta
\\
{{969576920701574621683}\over {7263766125000000}} &
b''b'b\delta^{(5)}\delta^{(3)}\delta''\delta'\delta\alpha &
{{969576920701574621683}\over {4358259675000000}} &
b^{(3)}b'b\delta^{(4)}\delta^{(3)}\delta''\delta'\delta\alpha
\\
{{-6922466437361968939693}\over {8150511600000000}} &
Tb''b'b\delta^{(4)}\delta^{(3)}\delta''\delta' &
{{1231279142989406788161469}\over {3137946966000000000}} &
Tb''b'b\delta^{(5)}\delta^{(3)}\delta''\delta
\\
{{-45875861358917235053603}\over {784486741500000000}} &
Tb''b'b\delta^{(5)}\delta^{(4)}\delta'\delta &
{{1354999687164657192816391}\over {3137946966000000000}} &
Tb''b'b\delta^{(6)}\delta^{(3)}\delta'\delta
\\
{{6799671218996227038570623}\over {21965628762000000000}} &
Tb''b'b\delta^{(7)}\delta''\delta'\delta &
{{36668428190742036245791}\over {209196464400000000}} &
Tb^{(3)}b'b\delta^{(4)}\delta^{(3)}\delta''\delta
\\
{{4550158381959521999637983}\over {4706920449000000000}} &
Tb^{(3)}b'b\delta^{(5)}\delta^{(3)}\delta'\delta &
{{452635413724118337879361}\over {448278138000000000}} &
Tb^{(3)}b'b\delta^{(6)}\delta''\delta'\delta
\\
{{285551042647987065597089}\over {627589393200000000}} &
Tb^{(3)}b''b\delta^{(4)}\delta^{(3)}\delta'\delta &
{{18674001411647853483407}\over {13584186000000000}} &
Tb^{(3)}b''b\delta^{(5)}\delta''\delta'\delta
\\
{{712530383295049840843673}\over {627589393200000000}} &
Tb^{(3)}b''b'\delta^{(4)}\delta''\delta'\delta &
{{11010843270748785177961}\over {11622025800000000}} &
Tb^{(4)}b'b\delta^{(4)}\delta^{(3)}\delta'\delta
\\
{{1062809816042147714999159}\over {784486741500000000}} &
Tb^{(4)}b'b\delta^{(5)}\delta''\delta'\delta &
{{1047526394937532878927871}\over {627589393200000000}} &
Tb^{(4)}b''b\delta^{(4)}\delta''\delta'\delta
\\
{{8926724537060433230933}\over {7844867415000000}} &
Tb^{(4)}b''b'\delta^{(3)}\delta''\delta'\delta &
{{5667634504872537413983}\over {15689734830000000}} &
Tb^{(4)}b^{(3)}b\delta^{(3)}\delta''\delta'\delta
\\
{{88573709328934131972631}\over {78448674150000000}} &
Tb^{(5)}b'b\delta^{(4)}\delta''\delta'\delta &
{{172495372529594372064287}\over {196121685375000000}} &
Tb^{(5)}b''b\delta^{(3)}\delta''\delta'\delta
\\
{{64636870642985996983147}\over {156897348300000000}} &
Tb^{(6)}b'b\delta^{(3)}\delta''\delta'\delta &
{{23733662073521791896167}\over {313794696600000000}} &
T^2b''b'b\delta^{(4)}\delta^{(3)}\delta'\delta
\\
{{-4812734794405584883421}\over {28526790600000000}} &
T^2b''b'b\delta^{(5)}\delta''\delta'\delta &
{{-2157272953547455133}\over {25357147200000}} &
T^2b^{(3)}b'b\delta^{(4)}\delta''\delta'\delta
\\
{{11028949252615013882357}\over {78448674150000000}} &
T^2b^{(3)}b''b\delta^{(3)}\delta''\delta'\delta &
{{-11617696545038738995243}\over {156897348300000000}} &
T^2b^{(4)}b'b\delta^{(3)}\delta''\delta'\delta
\\
{{1397336113203162961517}\over {3486607740000000}} &
T''Tb''b'b\delta^{(3)}\delta''\delta'\delta &
{{-5041067164155497640971}\over {78448674150000000}} &
T^3b''b'b\delta^{(3)}\delta''\delta'\delta .
\end{array}
\nonumber
\end{eqnarray}

Finally the 4 ghost number 5 terms are:
\begin{eqnarray}
\begin{array}{cl}
{{-3466641462380311323335551109341621}\over
  {1286130334232246580000000000}} &
b^{(3)}b''b'b\delta^{(5)}\delta^{(4)}\delta''\delta'\delta
\\
{{28623897568661550920687955013943}\over
  {13919159461387950000000000}} &
b^{(3)}b''b'b\delta^{(6)}\delta^{(3)}\delta''\delta'\delta
\\
{{39554725192676111386715222685079}\over {114833065556450587500000000}} &
b^{(5)}b''b'b\delta^{(4)}\delta^{(3)}\delta''\delta'\delta
\\
{{539380395627853044066179409239}\over {80192688254910000000000}} &
Tb^{(3)}b''b'b\delta^{(4)}\delta^{(3)}\delta''\delta'\delta .
\end{array}
\nonumber
\end{eqnarray}

\ack{I would like to thank Prof. C. Pope for reading this paper and for
discussions. Thanks also go to Prof. S. Randjbar-Daemi for inviting
me to visit ICTP where part of this work is done. This work is also
supported in part by Pan-Deng-Ji-Hua special project and
Chinese Center for Advanced Science and Technology (CCAST).}


\begin{thebibliography}{9}
\bibitem{zam}{A. B. Zamolodchikov, Theor. Math. Phys. 65 (1985) 1205.}
\bibitem{review}{For review see for example:
P. Bouwknegt and K. Schoutens, Phys. Reps. 223 (1993) 183.}
\bibitem{boot}{J. M. Figueroa-O'Farrill and J. Schrans, Phys. Lett.
B 245 (1990) 421.}
\bibitem{bonn}{R. Blumenhagen, M. Flohr, A. Kliem, W. Nahm,
A. Recknagel and R.  Varnhagen, Nucl. Phys. B 361 (1991) 255. }
\bibitem{wat}{H. G. Kausch and G. M. T. Watts, Nucl. Phys. B 354
(1991) 740. }
\bibitem{popea}{H. Lu, C. N. Pope and X. J. Wang,
Int. J. Mod. Phys. A 9 (1994) 1527. }
\bibitem{zhua}{C. -J. Zhu, Coments on the construction of
BRST charge for $W$ algebras, preprint in preparation}
\bibitem{zhub}{C. -J. Zhu, Nucl. Phys. B 418 (1994) 379. }
\bibitem{zhuc}{C. -J. Zhu, Phys. Lett.  B 316 (1993) 264. }
\bibitem{min}{J. Thierry-Mieg, Phys. Lett. B 197 (1987) 368. }
\bibitem{peter}{K. Schoutens, A. Sevrin and P. van Nieuwenhuizen,
Comm. Math. Phys. 124 (1989) 87.}
\bibitem{horn}{K. Hornfeck, Phys. Lett. B 315 (1993) 287.}
\bibitem{thie}{K. Thielemans, Intern. J. Mod. Phys. C 2 (1991) 787.}
\end{thebibliography}
\end{document}